\documentclass[longauth]{aa}  
\usepackage{xcolor}
\usepackage{txfonts}
\usepackage{hyperref}
\usepackage{amssymb}
\usepackage{graphicx}
\usepackage{amsmath,bm}
\usepackage{orcidlink}
\usepackage{pgfplots, pgfplotstable}
\usepackage{siunitx}
\usepackage{pict2e}
\usepackage{multirow}
\usepackage{dsfont}
\usepackage{appendix}
\usepackage{ulem}

\def\bl{B_{\mathrm{los}}}
\def\SIGP{\sigma_{\mathrm{\Phi}}}

\begin{document} 
\title{Direct assessment of SDO/HMI helioseismology of active regions on the Sun's far~side using  SO/PHI magnetograms} 
\titlerunning{Direct assessment of far-side helioseismology using  SO/PHI magnetograms} 
\author{
D. Yang\inst{1}\orcidlink{0000-0001-7570-1299}\thanks{\hbox{Corresponding author: Dan Yang}; \hbox{\email{yangd@mps.mpg.de}}}
   \and L. Gizon\inst{1,2,3}  \orcidlink{0000-0001-7696-8665}
   \and H. Barucq\inst{4} \and
   J.~Hirzberger\inst{1} \and
   D.~Orozco~Su\' arez\inst{5} \orcidlink{0000-0001-8829-1938} \and
   K.~Albert\inst{1} \orcidlink{0000-0002-3776-9548} \and
   N. Albelo~Jorge\inst{1} \and
   T.~Appourchaux\inst{6} \orcidlink{0000-0002-1790-1951} \and 
   A.~Alvarez-Herrero\inst{7} \orcidlink{0000-0001-9228-3412} \and
   J.~Blanco Rodr\'\i guez\inst{8}  \orcidlink{0000-0002-2055-441X} \and
   A.~Gandorfer\inst{1}  \orcidlink{0000-0002-9972-9840} \and
   D.~Germerott\inst{1} \and
   L.~Guerrero\inst{1} \and
   P.~Gutierrez-Marques\inst{1} \orcidlink{0000-0003-2797-0392} \and
   F. Kahil\inst{1} \orcidlink{0000-0002-4796-9527}  \and
   M.~Kolleck\inst{1} \and
   S.K.~Solanki\inst{1} \orcidlink{0000-0002-3418-8449} \and
   J.C.~del~Toro~Iniesta\inst{5} \orcidlink{0000-0002-3387-026X} \and
   R.~Volkmer\inst{9} \and
   J.~Woch\inst{1} \orcidlink{0000-0001-5833-3738} \and 
   I.~P\' erez-Grande\inst{10}\orcidlink{0000-0002-7145-2835} \and 
   E.~Sanchis~Kilders\inst{8} \orcidlink{0000-0002-4208-3575} \and
   M.~Balaguer~Jiménez\inst{5} \orcidlink{0000-0003-4738-7727} \and
   L.R.~Bellot~Rubio\inst{5} \orcidlink{0000-0001-8669-8857} \and
   D.~Calchetti\inst{1} \orcidlink{0000-0003-2755-5295} \and
   M.~Carmona\inst{11} \orcidlink{0000-0001-8019-2476} \and
   W.~Deutsch\inst{1} \and
   A.~Feller\inst{1} \and
   G.~Fernandez-Rico\inst{1,10} \orcidlink{0000-0002-4792-1144} \and
   A.~Fern\' andez-Medina\inst{7} \orcidlink{0000-0002-1232-4315} \and
   P.~Garc\'\i a~Parejo\inst{7} \orcidlink{0000-0003-1556-9411} \and 
   J.L.~Gasent~Blesa\inst{8} \orcidlink{0000-0002-1225-4177} \and 
   B.~Grauf\inst{1} \and 
   K.~Heerlein\inst{1} \and
   A.~Korpi-Lagg\inst{1} \orcidlink{0000-0003-1459-7074} \and
   T.~Lange\inst{12} \and  
   A.~L\' opez Jim\' enez\inst{5} \and 
   T.~Maue\inst{9, 13} \and 
   R.~Meller\inst{1} \and
   A.~Moreno Vacas\inst{5} \orcidlink{0000-0002-7336-0926} \and
   R.~M\" uller\inst{1} \and
   E.~Nakai\inst{9} \and 
   W.~Schmidt\inst{9} \and
   J.~Schou\inst{1} \orcidlink{0000-0002-2391-6156} \and
   U.~Sch\" uhle \inst{1} \orcidlink{0000-0001-6060-9078} \and
   J.~Sinjan \inst{1} \orcidlink{0000-0002-5387-636X} \and
   J.~Staub\inst{1} \orcidlink{0000-0001-9358-5834} \and
   H.~Strecker \inst{5} \orcidlink{0000-0003-1483-4535} \and 
   I.~Torralbo\inst{10} \orcidlink{0000-0001-9272-6439} \and 
   G.~Valori\inst{1} \orcidlink{0000-0001-7809-0067}
}

\institute{Max-Planck-Institut f\"ur Sonnensystemforschung,  Justus-von-Liebig-Weg 3, 37077 G{\"o}ttingen,  Germany \\ \email{yangd@mps.mpg.de} \\ \email{gizon@mps.mpg.de}
        \and
    Institut f\"ur Astrophysik, Georg-August-Universit{\"a}t G\"ottingen, Friedrich-Hund-Platz 1, 37077 G{\"o}ttingen, Germany   
       \and 
          Center for Space Science, NYUAD Institute, New York University Abu Dhabi, PO Box 129188, Abu Dhabi, UAE 
         \and
         Makutu, Inria, TotalEnergies, University of Pau, 64000 Pau, France
         \and       
         Instituto de Astrofísica de Andalucía (IAA-CSIC), Apartado de Correos 3004,
         18080 Granada, Spain 
         \and
         Univ. Paris-Sud, Institut d’Astrophysique Spatiale, UMR 8617,
         CNRS, B\^ atiment 121, 91405 Orsay Cedex, France
         \and
         Instituto Nacional de T\'ecnica Aeroespacial, Carretera de
         Ajalvir, km~4, 28850 Torrej\'on de Ardoz, Spain
         \and
         Universitat de Val\`encia, Catedr\'atico Jos\'e Beltr\'an 2, 46980 Paterna-Valencia, Spain
         \and
         Leibniz-Institut für Sonnenphysik, Sch\"oneckstr. 6, 79104 Freiburg, Germany
         \and
         Instituto Universitario ``Ignacio da Riva'', Universidad Polit\'ecnica de Madrid, IDR/UPM, Plaza Cardenal Cisneros 3, 28040 Madrid, Spain
         \and
         University of Barcelona, Department of Electronics, Carrer de Mart\'\i\ i Franqu\`es, 1--11, 08028 Barcelona, Spain
         \and
         Institut f\"ur Datentechnik und Kommunikationsnetze der TU
         Braunschweig, Hans-Sommer-Str.~66, 38106 Braunschweig,
         Germany
         \and
         Fraunhofer Institute for High-Speed Dynamics,
         Ernst-Mach-Institut, EMI, Ernst-Zermelo-Str. 4, 79104 Freiburg, Germany
          }
              
\date{Received \today ;  Accepted <date>}

\abstract
{Earth-side observations of solar p modes can be used to image and monitor magnetic activity on the Sun's far side. Here we use magnetograms of the far side obtained by the Polarimetric and Helioseismic Imager (PHI) onboard Solar Orbiter (SO) to directly assess -- for the first time -- the validity of far-side helioseismic holography. 
}
{We wish to co-locate the positions of active regions in helioseismic images and magnetograms, and to calibrate the helioseismic measurements in terms of magnetic field strength.
}
{We identify three magnetograms on 18 November 2020, 3 October 2021, and 3 February 2022 displaying a total of six active regions on the far side. The first two dates are from SO's cruise phase, the third from the  beginning of the nominal operation phase.
We compute  contemporaneous seismic phase maps for these three dates using helioseismic holography applied to time series of Dopplergrams from the Helioseismic and Magnetic Imager (HMI) on the Solar Dynamics Observatory (SDO). 
}
{
Among the six active regions seen in SO/PHI magnetograms, five active regions are identified on the seismic maps  at almost the same positions as on the magnetograms. One region is too weak to be detected above the seismic noise. To calibrate the seismic maps, we fit a linear relationship between the seismic phase shifts and the unsigned line-of-sight magnetic field averaged over the active region areas extracted from the SO/PHI magnetograms. 
}
{SO/PHI provides the strongest evidence so far that helioseismic imaging provides reliable information about active regions on the far side, including their positions,  areas,  and  mean unsigned magnetic field.} 


\authorrunning{D.~Yang, L.~Gizon, H.~Barucq, et al.}

\maketitle


\section{Introduction}
The solar surface provides important boundary conditions for the extrapolation of the structure and evolution of the global magnetic field into the heliosphere, where it interacts with the Earth's magnetosphere  \citep[see, e.g., review by][ and references therein]{Owens2013}. 
Half of the solar surface is invisible from an Earth vantage point, which leads to serious uncertainties when  modeling the  heliospheric field and limits the accuracy of space weather forecasts \citep{ARG13,Cash2015,WAL2022,Jain2022}. This missing part of the solar surface, often referred to as the Sun's far side, is however  accessible to spacecraft with large Earth\,--\,Sun\,--\,S/C angles, such as the {Solar Terrestrial Relations Observatory (STEREO)} and Solar Orbiter.    
The two STEREO spacecraft see the solar corona in the {extreme ultraviolet (EUV)} spectrum \citep{Howard2008}, while Solar Orbiter observes the corona as well as all the photospheric observables: intensity, Doppler velocity, and vector magnetic field \citep{MUL20,SOL20}.

\begin{figure*}[!htb]
\begin{center}
\includegraphics[width=\linewidth]{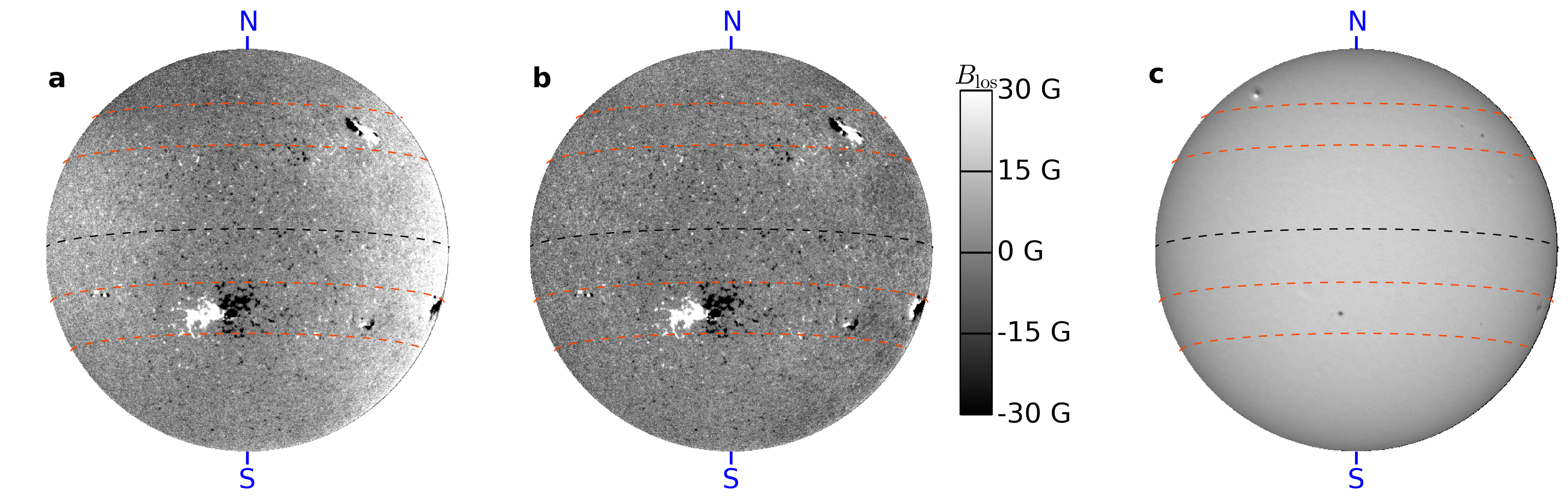} 
\caption{SO/PHI-FDT observations of the Sun's far side  on 18 November 2020 during Solar Orbiter's   cruise phase.
\textit{Panel a}: Line-of-sight magnetic field, $B_\mathrm{los}$.  
\textit{Panel b}: Corrected $B_\mathrm{los}$ after removing  the large-scale systematics by fitting a low-order polynomial  (see Sect.~\ref{sec.obs_phi}).
\textit{Panel c}: Continuum intensity, $I$. 
In each panel, the black dashed curve marks the solar equator and the orange dash curves highlight the latitude bands $25^\circ \le \lambda \le 40^\circ$ in the north and $-30 ^\circ \le \lambda \le -15^\circ$ in the south. 
} \label{fig.fdt_ccd}
\end{center}
\end{figure*}

\begin{table*}[htbp]
\medskip
\small
\begin{center}
\begin{tabular}{ccccccccccc}
\hline \hline
Index & Date & Time & CR & Latitude & Longitude   & { NOAA \#  } & Area   &   $\left\langle|\bl|\right\rangle_\mathrm{AR} $ & $\left\langle \Phi\right\rangle_\mathrm{AR}$  \\
 & & UTC & &  &  &     &  $10^4$ Mm$^2$ &   G & deg  \\
\hline

1& 18 Nov. 2020 & 02:40 &  2237 &$-18.0^\circ$ & $335.1^\circ$& 12786 (CR+1) & $1.63^\dagger$     &  $61^\dagger$ & $-14.8^\dagger$\\
2 &18 Nov. 2020 & 02:40& 2237 &$31.4^\circ$ & $302.1^\circ$ & 12787 (CR+1)  & $1.17$      &$45$ & $-11.3$\\
3& 18 Nov. 2020 &02:40 &2237 & $-26.3^\circ$ & $302.0^\circ$ & 12789 (CR+1) & $0.05$  &  $31$& $0.7^*$  \\
4& 18 Nov. 2020 & 02:40 & 2237 & $-23.5^\circ$ & $251.8^\circ$ & 12790 (CR+1) & $2.32$   &  $47$ & $-6.2$\\
5& 3 Oct. 2021 &01:30 &2249&$16.3^\circ$ & $155.0^\circ$& 12882 & $1.30$    &  $47$ & $-6.2$\\
6& 2 Feb. 2022 &23:30 &2253 &$23.8^\circ$  &$339.1^\circ$ &  12941 & $1.73$    & $61$&  $-8.8$\\ 
\hline
\multicolumn{10}{l} {$^\dagger$ Average quantities are computed only over the part of active region NOAA~12786 which is in SO/PHI's field of view.}\\
\multicolumn{10}{l} {$^*$ Below the level of seismic noise in the quiet Sun, $\SIGP=2.6^\circ$.}\\

\end{tabular}

\medskip
\caption{List of far-side active regions seen by SO/PHI. Each region is characterized by the date and time of SO/PHI observations, the coordinates of the region in the appropriate Carrington Rotation (CR) frame, and by its NOAA number when it can be connected to a region on the Earth side. The active region area, unsigned magnetic field ($|\bl|$), and helioseismic phase shifts ($\Phi$ from HMI) are averaged over the active region area determined by SO/PHI and outlined by the red contours in Figs.~\ref{fig.compare2d} and \ref{fig.compare2d_rest}.} \label{tab:ar_info}. 
\end{center}
\end{table*}

The Sun's far side is also accessible to helioseismic holography \citep[see, e.g.,][]{LIN00b, BRA02}, which provides information about magnetic activity in the solar near-surface layers. The input data for helioseismology are observations of solar acoustic oscillations from either the ground-based Global Oscillations Network Group  \citep[GONG,][]{Harvey1996} or the Solar Dynamics Observatory \citep[SDO,][]{Pesnell2012} in  geosynchronous orbit. 
This technique has been partially validated using far-side images of the chromosphere from STEREO/EUVI \citep{LIE12, Liewer2014,LIE17} or photospheric magnetograms that are shifted in time to wait for  active regions to come into Earth's view \citep{GON07}. In addition to helioseismic holography, time--distance helioseismology has also been used to  detect active regions on the far side and has been partially validated using STEREO chromospheric images \citep[e.g.,][]{ZHA07,ZHA19}.
Such a validation can only be partial since, in the first case, coronal images are not perfectly representative of the photospheric magnetic field and, in the second case, active regions evolve over daily and weekly time scales.
Thanks to recent advances in computational helioseismology \citep{GIZ17}, the detection rate of the active regions on the Sun's far side has improved significantly: STEREO data indicate  that most of the medium-sized active regions can be detected with confidence with helioseismic holography \citep{Yang2022}.

In order to be included  in models of the heliospheric field or space weather, the seismic measurements  need to be calibrated with respect to the photospheric magnetic field \citep[see a preliminary attempt by][]{ARG13}. 
Such calibrations have been considered on the front side using simultaneous SOHO/MDI line-of-sight magnetograms  \citep{Braun2001, Lindsey2005a, Lindsey2005b} and on the far side using (two-week delayed) GONG synoptic magnetograms   \citep{GON07} and STEREO-based proxies of the unsigned photospheric magnetic field \citep{Chen2022}. However, the optimal observations needed to calibrate far-side helioseismology  are contemporaneous images of the photospheric magnetic field, which were not available until the advent of Solar Orbiter (SO). The Polarimetric and Helioseismic Imager (PHI) on-onboard SO provided line-of-sight magnetograms of the far side during the cruise and the early science phase of the  mission \citep{SOL20}. 
In this paper, we aim to use these direct magnetic field measurements to assess and calibrate seismic  images on the Sun's far side using the recent improvements in helioseismic holography reported by \citet{Yang2022}.

\section{Observations}
\subsection{SO/PHI magnetograms  showing far side active regions}
\label{sec.obs_phi}

\begin{figure*}[!htb]
\begin{center}
\includegraphics[width=\linewidth]{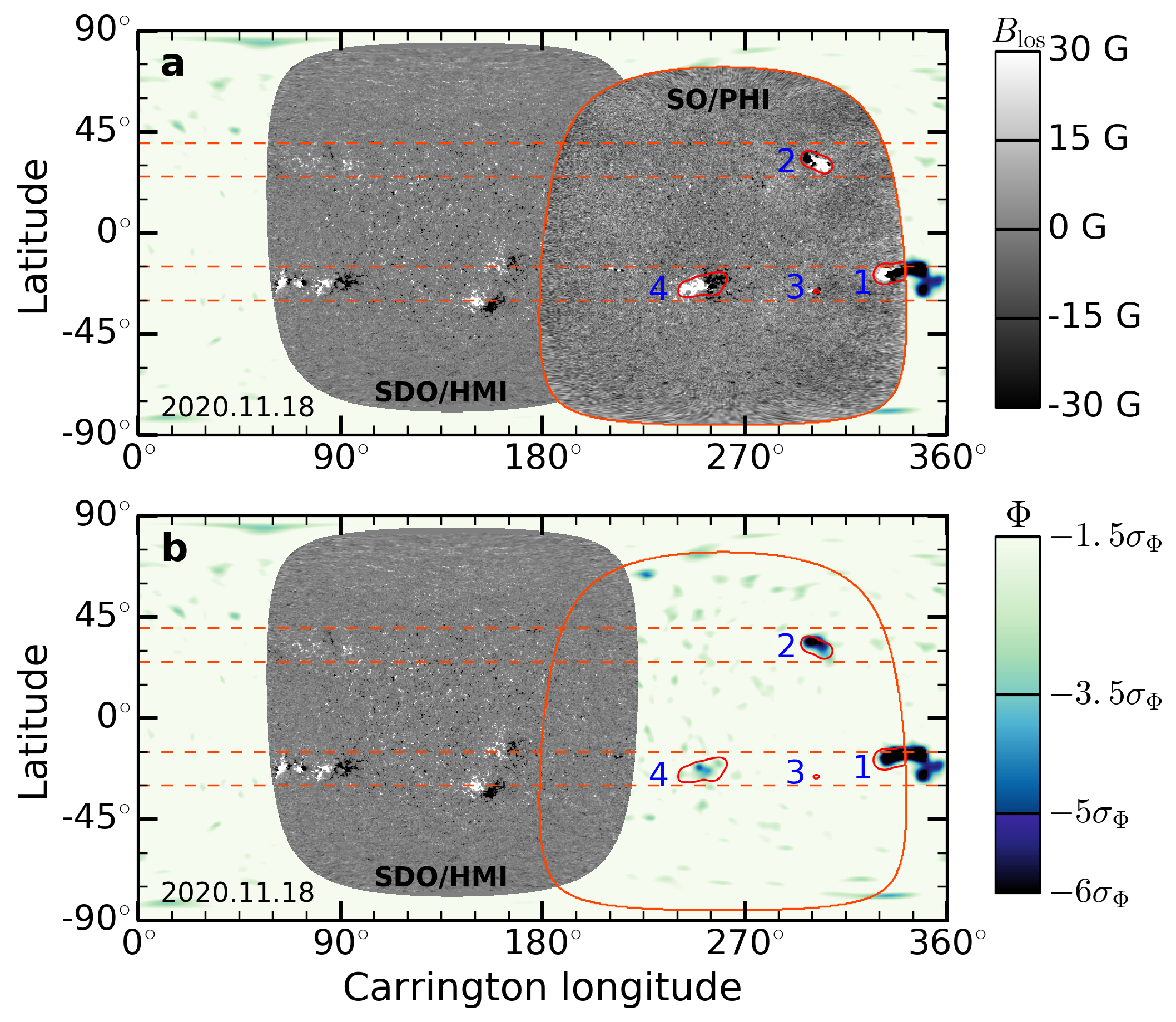} 
\caption{Magnetic activity on the entire solar surface on 18 November 2020 during Carrington Rotation CR 2237.
\textit{Panel a}: The SO/PHI-FDT magnetogram covers a large fraction of the far side, while a 3-day averaged SDO/HMI magnetogram shows magnetic activity on the near side.
Four active regions identified on the far side by SO/PHI are outlined by red contours (See Table~\ref{tab:ar_info}). 
\textit{Panel b}: The green/blue shades show the helioseismic phase $\Phi$ on the far side, deduced from acoustic oscillations observed on the near-side by SDO/HMI over 79 hours during 17--19 November. The seismic phase is shown  over a range corresponding to $1.5$--$6$ times the standard deviation of the noise in the quiet Sun ($\SIGP = 2.6^\circ$). }  \label{fig.compare2d}
\end{center}
\end{figure*}

We use the magnetograms from the SO/PHI-Full {Disk} Telescope (FDT), which measures the Stokes parameters of  the Fe I  6173~\si{\angstrom} spectral line at 6 wavelength positions \citep{SOL20}. 
During the cruise phase, SO/PHI was only switched on occasionally within the so-called ``remote-sensing checkout windows'' during which it provided at most a few magnetograms per day \citep{Zou2020}. Daily synoptic magnetograms covering several months will be made available relatively soon, however, this data set are still being processed and thus are not included in this work. A multi-view synoptic map for a single Carrington rotation was processed by \citet{Loeschl2023}.
The SO/PHI raw data can be reduced and processed either on-board \citep{Kinga2020}  or on the ground. The on-ground data processed includes additional corrections (fringe and ghost corrections).

The present work uses  SO/PHI-FDT magnetograms, each with $2048 \times 2048$ pixels but with varying spatial sampling of the solar disk due to a large range of Sun-S/C distances \citep[{highest resolution at disk center of $\approx 730$~km reached at  $0.28$~au,} see][]{SOL20}. In total, three magnetograms are analyzed, corresponding to the specific times when SO/PHI saw active regions on the far side. These data are from the cruise phase on 18 November 2020 and 3 October 2021 {(processed on-board)}, and from the nominal science phase on 3 February 2022 {(processed on-ground)}.
At the time of writing, the magnetograms have been reduced in a preliminary manner only and consequently are affected by large-scale imperfections across the field of view (different for each image). We estimate these systematics by fitting a low-order polynomial of the form $\sum_{i,j=0}^{3} a_{ij} x^i y^j$ to the data using a least squares method, where $x$ and $y$ are the pixel coordinates and the fit applies only to the pixels that are outside of active regions. Example raw and processed magnetograms (after subtraction of the large-scale systematics) are shown in Fig.~\ref{fig.fdt_ccd}.

\begin{figure*}[t]
\begin{center}
\includegraphics[width=\linewidth]{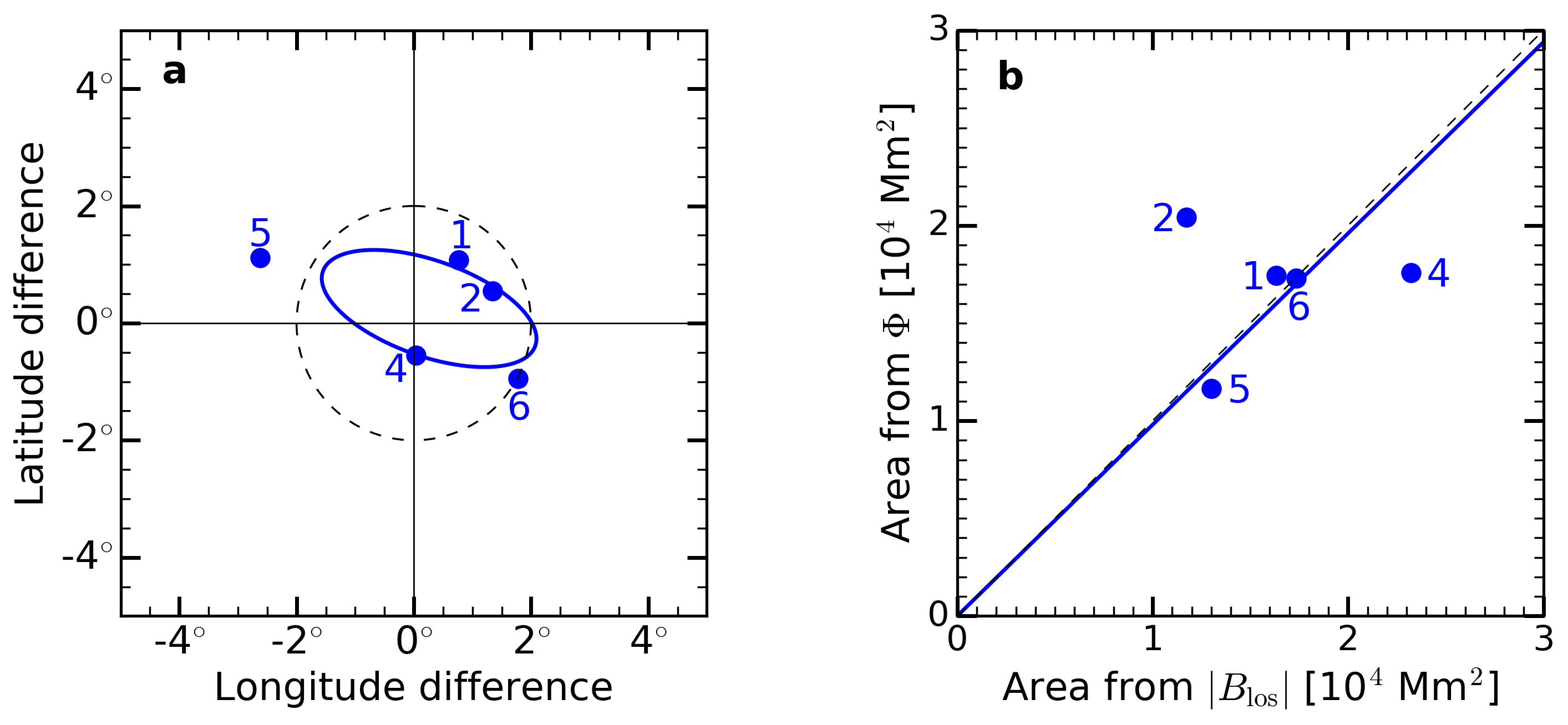}
\caption{
\textit{Panel a}: Positions of active regions 1, 2, 4, 5, and 6 deduced from the seismic maps, with respect to their actual positions deduced from the SO/PHI magnetograms.
The active region coordinates are computed using a center-of-gravity method, applied to each observable ($\Phi$ or $\bl$) within the region areas.
For the SO/PHI magnetograms, the active region areas correspond to the red contours in Figs.~\ref{fig.compare2d} and~\ref{fig.compare2d_rest} ($20$~G contours from the smoothed magnetograms). For the helioseismic data, a threshold of $2 \SIGP$ is used to determine the active region areas. 
 The blue ellipse shows the probability density function  of the data points at half maximum, assuming a bivariate normal distribution. 
The dashed circle with a diameter $4^\circ$ corresponds to the resolution limit of the seismic observations. 
\textit{Panel b}:
Active region areas inferred from helioseismology versus areas measured from the SO/PHI magnetograms. The linear fit gives a slope of $0.98$ (solid blue line). The dashed line is the 1:1 diagonal.
}\label{fig.position}
\end{center}
\end{figure*} 

\subsection{Helioseismic maps}
\label{sec.obs_seismic}
We apply far-side helioseismic holography to series of Dopplergrams from the Helioseismic and Magnetic Imager \citep[HMI,][]{SCH12} onboard SDO.
The data reduction follows exactly the procedure described in detail by   \citet{Yang2022}. Helioseismic holography measures the phase shift $\Phi$ between the ingression and the egression, i.e. between the forward- and the backward-propagated acoustic waves from the  observed surface to any target location on the Sun's far side. 
The acoustic waves propagate faster in magnetized regions on the surface, which will cause a negative shift in $\Phi$. The maps of $\Phi$ can be used to detect individual active regions on the far side and to  monitor their evolution \citep{Yang2022}.

Following \citet{Yang2022}, we use seismic phase maps averaged over 79 hours. These phase maps are given in Carrington coordinates (during CR 2237, 2249, and 2253) 
and are smoothed with a Gaussian kernel of full width $4^\circ$ to remove the  spatial scales smaller than half the  typical wavelength on the surface of the seismic waves used in the analysis.
The far side maps centered on 18 November 2020 (Fig.~\ref{fig.compare2d}b) and 3 October 2021 (Fig.~\ref{fig.compare2d_rest}b) were computed during periods when the duty cycle of  SDO/HMI Dopplergrams was almost 100~\%.   
The third phase map centered on 3 February 2022 was computed during the eclipse season of SDO (24 January--17 February 2022), a period of lower quality images and lower duty cycle  (94~\%, 92~\%, and 77~\% on February 2, 3, and 4 2022). The lower data rate on February 4 is mainly due to snow on the receiving antenna dish. This explains the higher seismic noise  for the 79-hr average  phase map centered on 3 February 2022 (Fig.~\ref{fig.compare2d_rest}d).

\section{Direct assessment of helioseismic imaging}

Figure~\ref{fig.compare2d}a shows the SO/PHI line-of-sight magnetograms  on 18 November 2020, which includes several active regions on the far side marked by red contours. The other two magnetograms are shown in the appendix in Figs.~\ref{fig.compare2d_rest}a and ~\ref{fig.compare2d_rest}c. In total, we identified six far-side active regions in the collection of SO/PHI magnetograms that were available to us,  including five relatively big regions and a much smaller region. The parameters of these far-side active regions are  given in Table~\ref{tab:ar_info}.  Tracking these regions on the far side enables to {assign to each of them a} NOAA number as they rotate onto the front side.
Fig.~\ref{fig.compare2d}b, Fig.~\ref{fig.compare2d_rest}b, and Fig.~\ref{fig.compare2d_rest}d show the values of the seismic phase over a range corresponding to $1.5$--$6$ times the standard deviation of the noise in the quiet Sun ($\SIGP = 2.6^\circ$).  
The five largest active regions observed by SO/PHI are clearly visible in the contemporaneous helioseismic phase maps, at almost the same positions and with similar areas as seen on the magnetograms.

Figure~\ref{fig.position}a shows the positions of the active regions determined from the seismic maps with respect to their 
true positions measured from the SO/PHI magnetograms. These positions are obtained by computing the center of gravity of the data over the corresponding active region areas. 
For the SO/PHI data, the active region areas are defined as the areas with magnetic field above $20$~G from the unsigned magnetograms smoothed by a Gaussian kernel of width $4^\circ$.  For the helioseismic data, we first identify the active regions using a detection threshold of $-3.5 \SIGP$ \citep[as proposed by][see green curves in Fig.~\ref{fig.zoom}]{Yang2022} and then extend the active region areas using a lower threshold of $-2 \SIGP$ (see blue curves in Fig.~\ref{fig.zoom}). Only the data within SO/PHI's field of view are used in the analysis. As shown in Fig.~\ref{fig.position}a, all 5 active regions detected in the seismic maps are within a few degrees of their true positions. This remarkable result is consistent with  the  resolution limit of seismic imaging (half the typical acoustic wavelength at the surface corresponds to an angular distance of approximately $4^\circ$).

{Figure~\ref{fig.position}b shows that the helioseismically-determined areas of the active regions ($2 \SIGP$ threshold) are similar to the true areas determined from the SO/PHI magnetic field data. While this result is encouraging, a definitive statement would require the analysis of a much larger data set.
Active regions areas are useful to measure on far side, for example for solar irradiance computations (and predictions) from a variety of vantage points.}

These all demonstrate  that far-side helioseismic holography works well. The seismic phase shifts  associated with the active regions may then be empirically related to magnetic field strength measured by SO/PHI. For the small active region in the South in Fig.~\ref{fig.compare2d} 
 (active region 3), no excess signal is seen in the seismic phase maps. 

\begin{figure}
\begin{center}
\includegraphics[width=\linewidth]{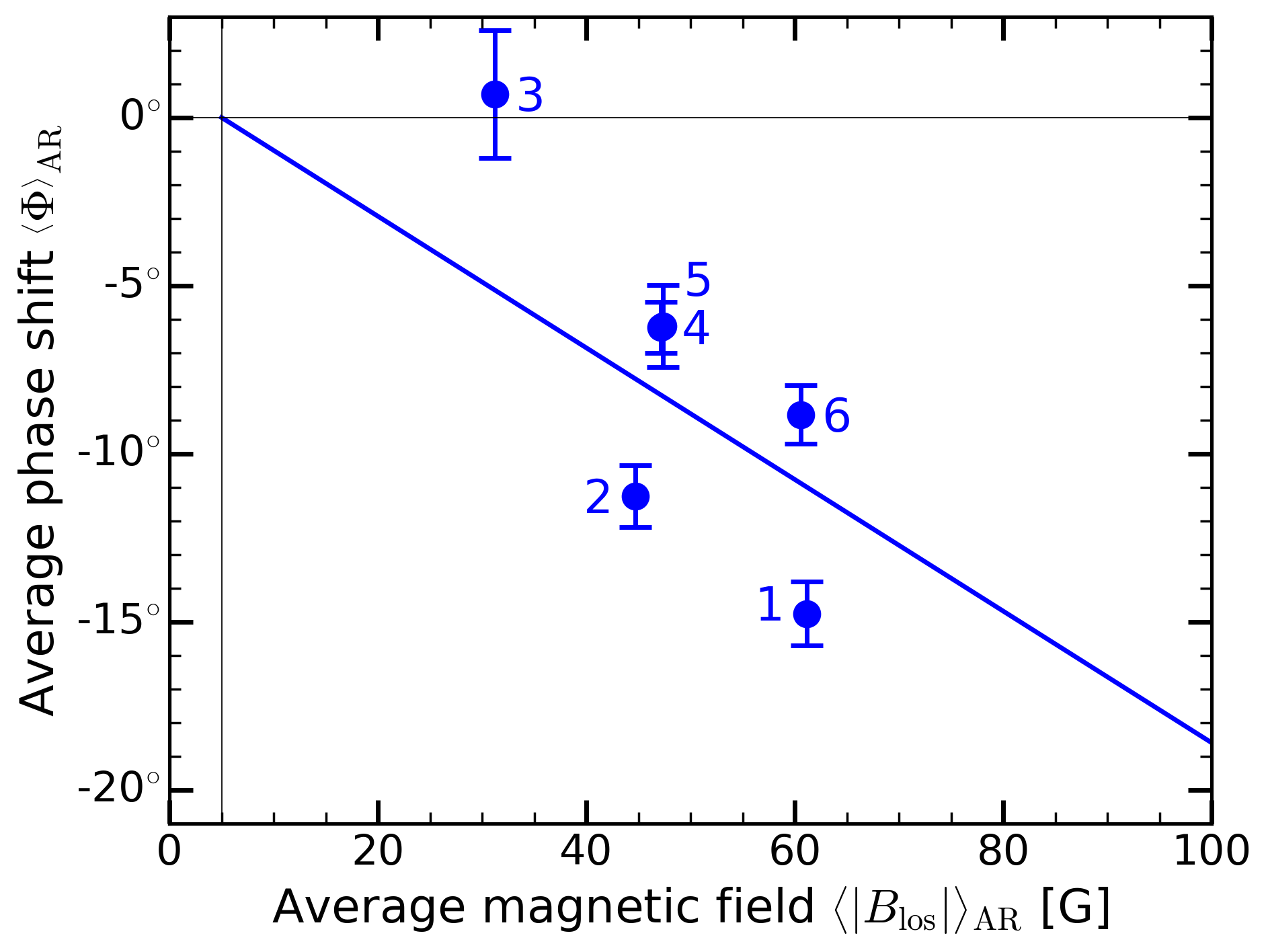}
\caption{
Helioseismic phase shift  $\Phi$ versus unsigned line-of-sight magnetic field $|\bl|$, both averaged over each of the six active regions on the far side  (blue dots). Active region areas are determined from SO/PHI magnetograms (red contours  in Figs.~\ref{fig.compare2d} and ~\ref{fig.compare2d_rest}). The numerical values are provided in  Table~\ref{tab:ar_info} and  the error bars refer to $\pm 1 \SIGP$ seismic noise. The blue solid line is a linear fit with slope  $-0.2\ \mathrm{deg/G}$ and intercept (5~G, $0^\circ$). 
}\label{fig.calibration} 
\end{center}
\end{figure}

\begin{figure*}
\begin{center}
\includegraphics[width=\linewidth]{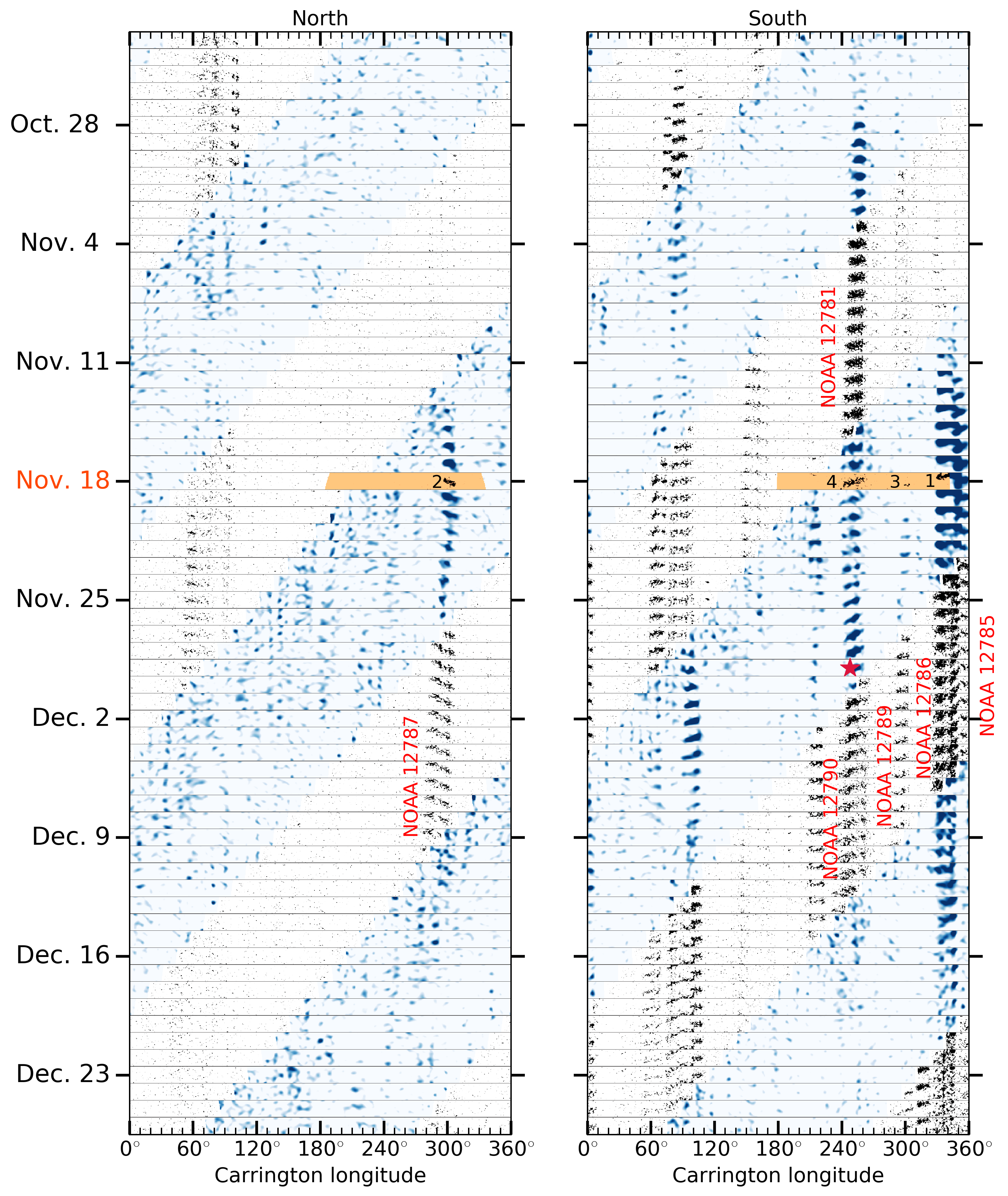}
\caption{Longitude–time diagrams based on subsets of maps such as Fig.~\ref{fig.compare2d}b. \textit{Left panel}: Latitude bands  in the northern hemisphere ($25^\circ \le \lambda \le 40^\circ$, see Fig.~\ref{fig.compare2d}) stacked in time to cover the period from 23 October to 26 December 2020. The far side shows calibrated phase maps, $B_\Phi$,  in the range $20$--$50$~G (blue shades). The near side shows the unsigned line-of-sight magnetic field $\bl$ from SDO/HMI (gray shades) in the same range.
The unsigned $\bl$ from SO/PHI-FDT for 18 November 2020 is shown in orange.  
\textit{Right panel}: Same as the left panel but for latitude bands in the southern hemisphere ($-30 ^\circ \le \lambda \le -15^\circ$, see Fig.~\ref{fig.compare2d}). The red star indicates the occurrence of an M~4.4 class flare near the limb on 29 November 2020, associated with NOAA~12790. 
} \label{fig.lat_time_plot}
\end{center}
\end{figure*}

\section{Preliminary calibration of seismic maps}
\label{sec.calibration}

We compute the values of $|\bl|$ and $\Phi$ averaged over the areas of the six active regions detected on the far side and outlined by the red contours (e.g., in Fig.~\ref{fig.compare2d}) extracted from the SO/PHI-FDT magnetograms.
These spatial averages over active region areas are denoted with angle brackets $\langle \cdot \rangle_{\rm AR}$.
As shown in Fig.~\ref{fig.calibration}, the data suggest a trend where $\langle\Phi\rangle_\mathrm{AR}$ decreases with $\langle|\bl|\rangle_\mathrm{AR}$ (i.e. increases in absolute value).

With only six data points, the exact functional relationship between the two quantities is difficult to determine. Once the zero point is fixed, we find that a linear fit is as good as a quadratic fit to describe the data  (reduced $\chi^2$ of $11.1$ versus $11.0$). 
For the linear fit we obtain
\begin{equation}
\left\langle\Phi\right \rangle_\mathrm{AR}
 = a \left(\left\langle|\bl|\right\rangle_\mathrm{AR} - 5~\textrm{G}\right), \qquad a = -0.2~ \textrm{deg/G},
 \label{eq.calibration}
\end{equation}
where the quiet Sun reference  $|\bl|\approx 5$~G is measured near solar minimum  \citep[see also][]{Lagg2022}. We note that the scatter of the data points in Fig.~\ref{fig.calibration} could potentially be reduced by taking into account the positions of active regions with respect to disk center in both maps -- we intend to study this possibility in the future when many more active regions will be available from the SO/PHI synoptic program.

Figure~\ref{fig.1dslice} shows plots of the seismic phase calibrated using the transformation
\begin{equation}
\Phi \rightarrow  B_\Phi  := \Phi / a +  5\ \textrm{G}.
\label{eq.relation}
\end{equation}
The quantity $B_\Phi$ can be understood as a seismic estimate of the unsigned magnetic field $|\bl|$. In Fig.~\ref{fig.1dslice}, 
both $B_\Phi$ and  $|\bl|$ were averaged over the latitude bands shown in Figs.~\ref{fig.compare2d}a and~\ref{fig.compare2d}b for  active region 2 (panel a) and active regions 1 and 4 (panel b)  on 18 November 2020. $B_\Phi$ matches reasonably well with  $|\bl|$ observed by SO/PHI for all three active regions in terms of position and shapes. With the proposed calibration, we find very similar amplitudes for $B_\Phi$ and $|\bl|$ for  active regions 1 and 4 (Fig.~\ref{fig.1dslice}b), while a factor of two difference is seen  for active region 2 (Fig.~\ref{fig.1dslice}a).

\section{Emergence and evolution of active regions}
With STEREO chromospheric images, \citet{Yang2022} showed  that seismic phase maps can be used to study the evolution of active regions on the far side. By applying the scaling relation Eq.~(\ref{eq.relation}) to the seismic phase maps and by combining them with front side magnetograms, we can follow the evolution of active regions over multiple rotations  in a consistent manner. Such a longitude-time plot combining $B_\Phi$ and $|\bl|$ is shown in Fig.~\ref{fig.lat_time_plot} for the period from 23 October to 26 December 2020, where active regions NOAA 12787 (in the north), 12781/90, and 12786 (in the south) can be followed. The magnetic proxy $B_\Phi$ is shown in the range $20$\,--\,$50$~G on the far side, while the Earth side shows $|\bl|$ from SDO/HMI in the same range. $|\bl|$ from SO/PHI-FDT (orange) is shown for the date of 18 November 2020. Figure~\ref{fig.lat_time_plot} gives a consistent view for the evolution of the active regions from their emergence until they disperse. In particular, we see clearly that NOAA 12781 and 12790 refer to the same active region. 
Active region NOAA 12781 emerges on 28 October 2020, grows until it comes into Earth's view, and then starts to disperse. This region later rotates on the far side, reappears on the near side as NOAA 12790, and then fades away with a total life of $\approx 45$ days. 
Thus far-side imaging (validated with SO/PHI) enables to measure the time of emergence of active regions and to track long-lived regions across multiple Carrington rotations. Another representation of the evolution of this active region is given by  Fig.~\ref{fig.evolution_1d_slice}.

\section{Conclusion }
We have shown in this work how the SO/PHI magnetograms are key to {validating} far-side helioseismic imaging. We find that the few active regions  identified on the far side  are located in seismic maps at almost the same positions and with similar areas as in the SO/PHI {line-of-sight} magnetograms. Furthermore, the seismic phase associated with an active region can be empirically related to the average unsigned magnetic field. 
This work gives the strongest evidence so far that seismic imaging -- as implemented by \citet{GIZ18} and \citet{Yang2022} --  provides reliable information about active regions on the Sun's far side. 
This calibration is not yet perfect since only the line-of-sight component of the field is available (at the time of writing) and we only considered six active regions in three magnetograms.
Once extended to the many more far-side active regions to be observed by SO/PHI, we will be able to improve the calibration of the seismic maps.
This will lead to an improved understanding of  the  emergence and evolution of active regions over the entire solar surface.

\begin{acknowledgements}
We thank an anonymous referee for their thoughtful comments. DY and LG acknowledge funding from ERC Synergy Grant WHOLE SUN (\#810218) and  Deutsche Forschungsgemeinschaft  (DFG) Collaborative Research Center SFB 1456 (\#432680300, project C04). LG acknowledges NYUAD Institute Grant G1502. HB acknowledges funding from the 2021-0048: Geothermica SEE4GEO of European project  and the associated team program ANTS of Inria. The HMI data are courtesy of SDO (NASA) and the HMI consortium.
Solar Orbiter is a space mission of international collaboration between ESA and NASA, operated by ESA.  We are grateful to the ESA SOC and MOC teams for their support.  The German contribution to SO/PHI is funded by the BMWi through DLR and by MPG central funds. The Spanish contribution is funded by AEI/MCIN/10.13039/501100011033/ (RTI2018-096886-C5, PID2021-125325OB-C5, PCI2022-135009-2) and ERDF “A way of making Europe”,  Center of Excellence Severo Ochoa Awards to IAA-CSIC (SEV-2017-0709, CEX2021-001131-S), and a Ramón y Cajal fellowship  to DOS. The French contribution is funded by CNES. 
\end{acknowledgements}

\bibliographystyle{aa}
\bibliography{biblio.bib}
 
\appendix
\section{Additional figures}
\begin{figure*}
\begin{center}
\includegraphics[width=\linewidth]{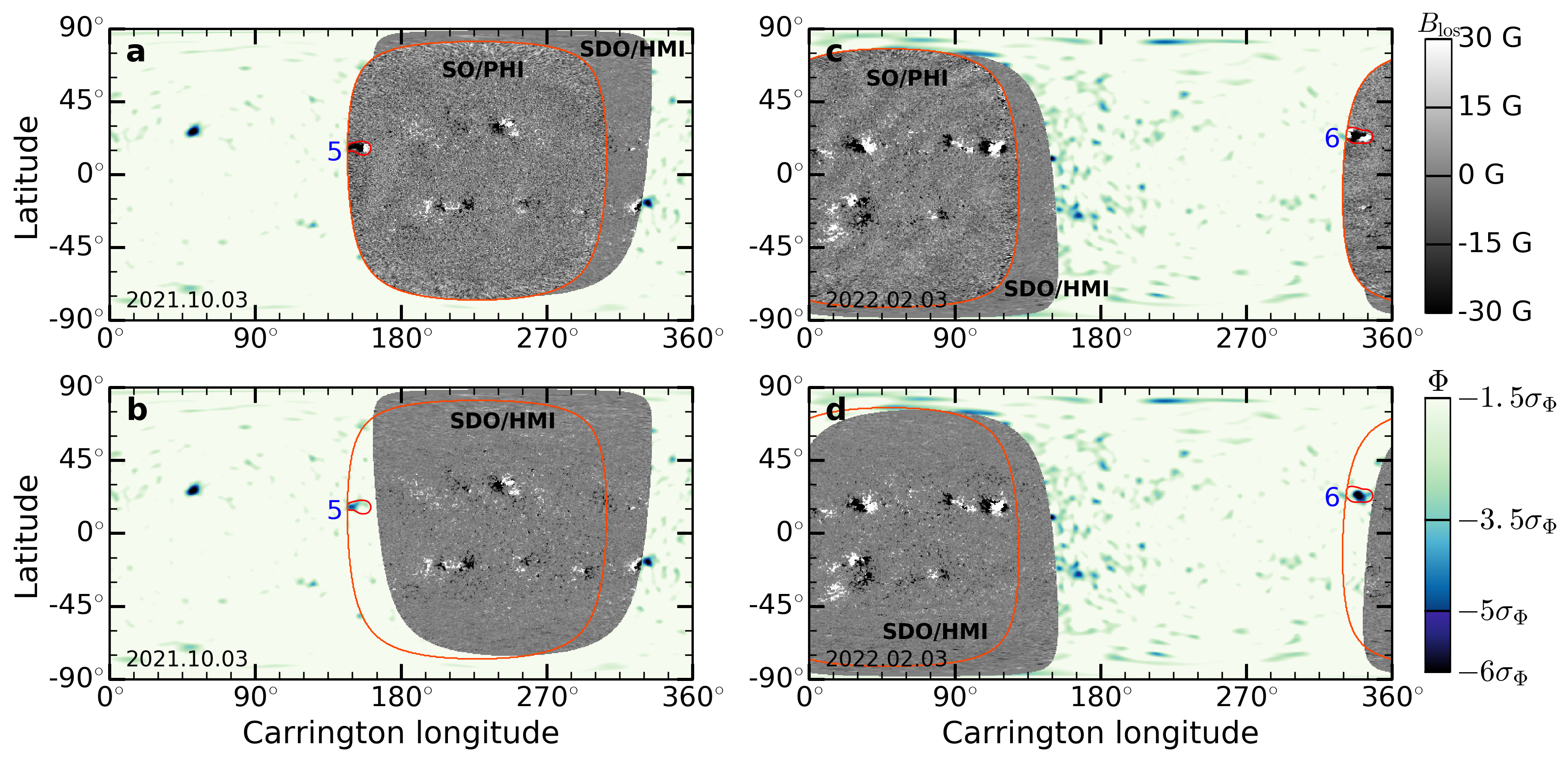}
\caption{Same as Fig.~\ref{fig.compare2d}, but for the other two active regions identified with  SO/PHI on the far side (red contours):  Active region 5 on 3 October 2021 (left panels) and active region 6 (right panels) on 3 February 2022. Both regions appear on the near side about one day later.  The higher seismic noise in the right panels is due to lower-quality SDO/HMI data (see main text).
} \label{fig.compare2d_rest}
\end{center}
\end{figure*} 

\begin{figure*}
\begin{center}
\includegraphics[width=\linewidth]{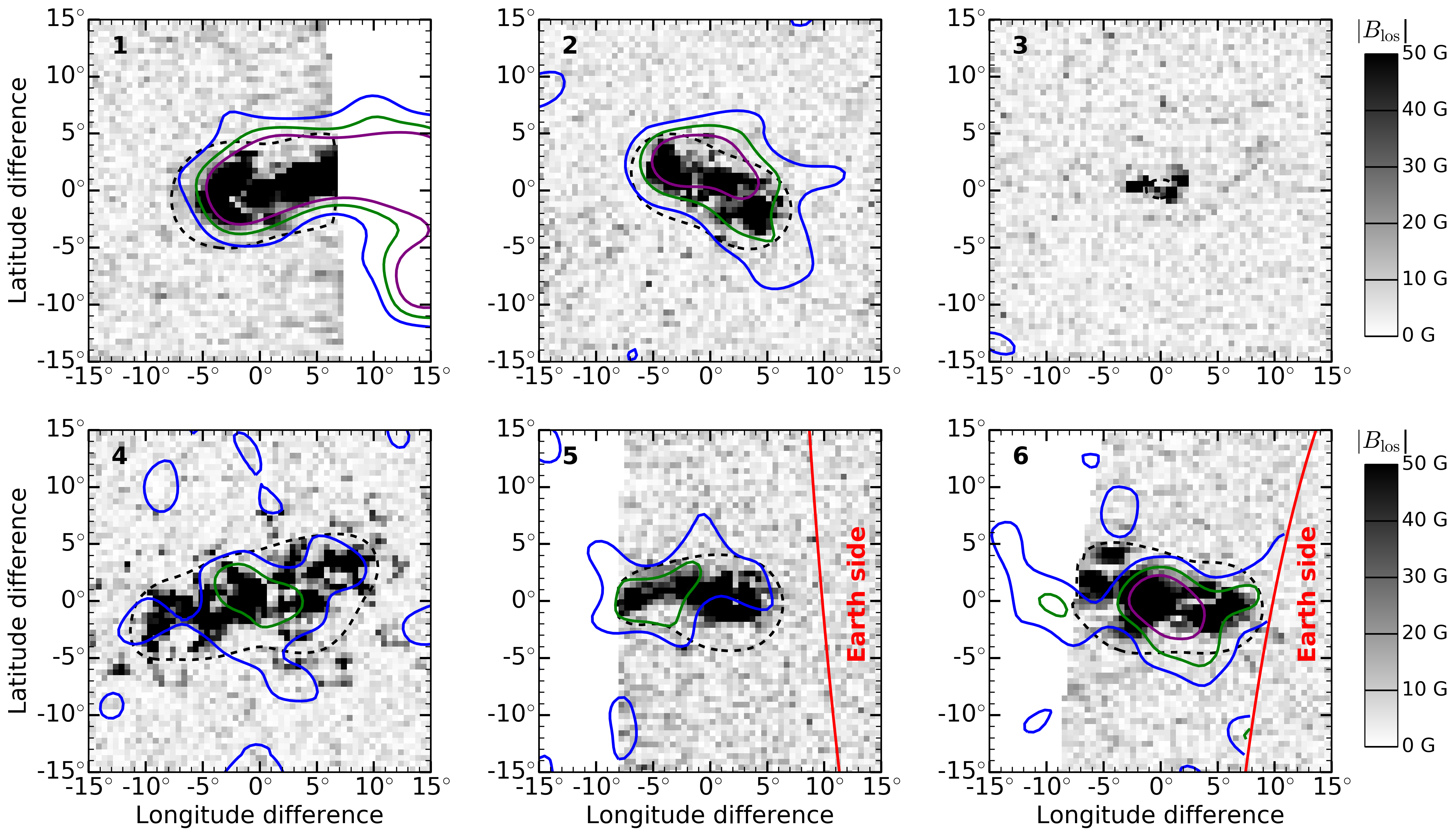}
\caption{Contour plots of  the helioseismic phase  overlaid on SO/PHI unsigned magnetograms (gray scale saturated at $50$~G) for far-side active regions 1 to 6 (see labels in top-left corners).  The black dashed contours correspond to $|\bl| = 20$ G after Gaussian smoothing. The blue, green, and purple contours correspond to $\Phi=-2 \SIGP$, $-3.5 \SIGP$, and $-5 \SIGP$ respectively.   The red curves  mark the boundaries between the far and Earth sides. Active region 3 is not detected by helioseismology.} \label{fig.zoom}
\end{center}
\end{figure*}

\begin{figure*}
\begin{center}
\includegraphics[width=\linewidth]{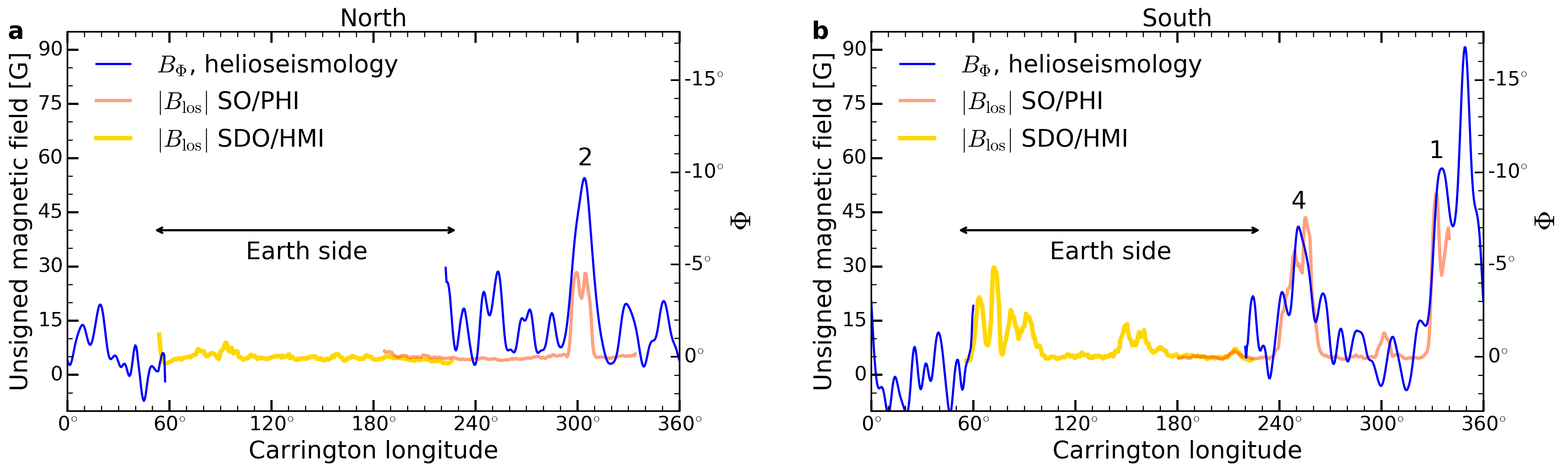}
\caption{
Plots of $B_\Phi$ from far-side helioseismology for 17--19 November 2020 (blue) and unsigned line-of-sight magnetic field $|\bl|$ on 18 November 2020 from SO/PHI (orange) and SDO/HMI (yellow). The data are averaged over the latitude band
$25^\circ \le \lambda \le 40^\circ$ in the north ({\it panel~a}) and over the latitude band $-30 ^\circ \le \lambda \le -15^\circ$ in the south ({\it panel~b}).
 Note that  $|\bl|$ is  smoothed in longitude with a running mean over $4^\circ$. The active regions 1, 2, and 4 are indicated on the plots (see Table~\ref{tab:ar_info}). Notice that the values shown here for active regions 1, 2, and 4 are not directly comparable to the values given in Fig.~\ref{fig.calibration} because we are averaging over latitude bands and not over active region areas.
 } \label{fig.1dslice}
\end{center}
\end{figure*} 

\begin{figure*}
\begin{center}
\includegraphics[width=0.68\linewidth]{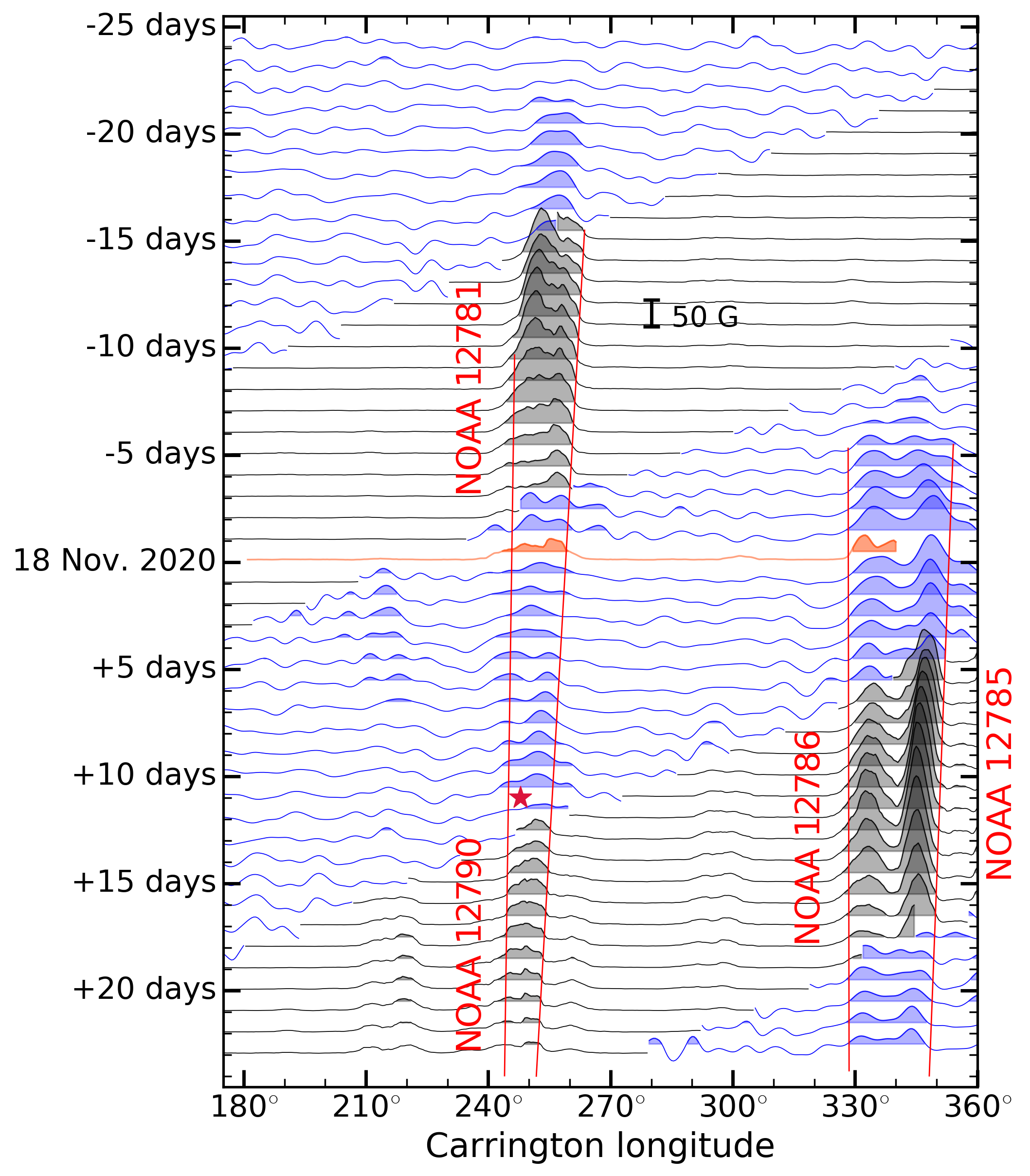}
\caption{
Average of the data from the right panel of Fig.~\ref{fig.lat_time_plot} over the latitudinal band $-30 ^\circ \le \lambda \le -15^\circ$ in the South. The averages of $B_\Phi$ are given by the blue curves (far side), the averages of $|\bl|$ by the black curves (near side), and the average of $|\bl|$ from SO/PHI for 18 November 2020 by the orange curve. The magnetic field data are further smoothed in longitude with a running mean over $4^\circ$. The filled areas below the curves highlight the values above $20$~G. The red star indicates the same flare (M~$4.4$) as shown in Fig.~\ref{fig.lat_time_plot}. 
 } \label{fig.evolution_1d_slice}
\end{center}
\end{figure*}

\end{document}